\documentclass[12pt]{article} 
\usepackage{amssymb}
\usepackage{graphicx}
\begin{document} 
\begin{center}
{\large \bf The critical regime of elastic scattering of protons at the LHC}

\vspace{0.5cm}                   

{\bf I.M. Dremin\footnote{E-mail: dremin@lpi.ru}}

\vspace{0.5cm}                       

         Lebedev Physical Institute, Moscow 119991, Russia\\

\end{center}

\begin{abstract}
It is shown that the darkness of the interaction region of protons is
governed by the ratio of the slope of the diffraction cone to the total
cross section. At LHC energies, it becomes completely absorptive at small
impact parameters. The lower limit of the ratio is determined. That imposes
some restrictions on its energy behavior. It is argued that the black disk
terminology should be replaced by the black torus.
\end{abstract}

The total cross section of colliding protons $\sigma _t$ depends on their 
energy. Another important experimental characteristics is the slope $B$
of the differential cross section of elastic scattering. Both of them increase 
with energy at high energies. Let us show that their ratio uniquely defines
the darkness (opacity) at the very center of the interaction region.

The differential cross section of elastic scattering $d\sigma /dt$ is 
related to the scattering amplitude $f(s,t)$ in a following way
\begin{equation}
\frac {d\sigma }{dt}=\vert f(s,t)\vert ^2.
\label{dsdt}
\end{equation}
Here $s=4E^2$, where $E$ is the energy in the center of mass system. The 
four-momentum transfer squared is
\begin{equation}
-t=2p^2(1-\cos \theta ) 
\label{trans}
\end{equation}
with $\theta $ denoting the scattering angle in the center of mass system
and $p$ the momentum. The amplitude $f$ is normalized at $t=0$
by the optical theorem such that
\begin{equation}
{\rm Im}f(s,0)=\sigma _t/\sqrt {16\pi}.
\label{opt}
\end{equation}
Note that the dimension of $f$ is GeV$^{-2}$.

It is known from experiment that protons mostly scatter at rather small
angles within the so-called diffraction cone. As a first approximation, it
can be described by the exponential shape with the slope $B$ such that
\begin{equation}
\frac {d\sigma }{dt}\propto e^{-B\vert t\vert }.
\label{expB}
\end{equation}

To define the geometry of the collision we must express these characteristics
in terms of the transverse distance between the centers of the colliding
protons called the impact parameter $b$. It is easily done by the 
Fourier-Bessel transform of the amplitude $f$ written as
\begin{equation}
i\Gamma (s,b)=\frac {1}{2\sqrt {\pi }}\int _0^{\infty}d\vert t\vert f(s,t)
J_0(b\sqrt {\vert t\vert }).
\label{gamm}
\end{equation}
Using the above formulae, one can write the dimensionless $\Gamma $ as
\begin{equation}
i\Gamma (s,b)=\frac {\sigma _t}{8\pi }\int _0^{\infty}d\vert t\vert 
e^{-B\vert t\vert /2 }(i+\rho (s,t))J_0(b\sqrt {\vert t\vert }).
\label{gam2}
\end{equation}
Here $\rho (s,t) = {\rm Re}f(s,t)/{\rm Im}f(s,t)$ and the diffraction cone
approximation (\ref{expB}) is inserted. Herefrom, one calculates
\begin{equation}
{\rm Re}\Gamma (s,b)=\frac {1}{Z}e^{-\frac {b^2}{2B}},
\label{rega}
\end{equation}
where $Z=4\pi B/\sigma _t$ is the variable used in the review paper 
\cite{ufnel}. This dependence on the impact parameter was used, in particular, 
in \cite{fsw}. 

The elastic scattering amplitude must satisfy the most general principle of
unitarity which states that the total probability of outcomes of any particle 
collision sums to 1 and reads
\begin{equation}
G(s,b)=2{\rm Re}\Gamma (s,b)-\vert \Gamma (s,b)\vert ^2.
\label{unit}
\end{equation}
The left-hand side called the overlap function describes the impact-parameter
profile of inelastic collisions of protons. It satisfies the inequalities
$0\leq G(s,b)\leq 1$ and determines how absoptive is the interaction region
depending on the impact parameter (with $G=1$ for full absorption).

It is known from experiment that the ratio $\rho (s,t)$ is very small at
$t=0$ and, at the beginning, we neglect it and get
\begin{equation}
G(s,b)= \frac {2}{Z}e^{-\frac {b^2}{2B}}-\frac {1}{Z^2}e^{-\frac {b^2}{B}}.
\label{ge}
\end{equation}
For central collisions with $b=0$ it gives
\begin{equation}
G(s,b=0)= \frac {2Z-1}{Z^2}.
\label{gZ}       
\end{equation}
Thus, the darkness of the central region is fully determined by the ratio $Z$. 
It becomes completely absorptive only at $Z=1$ and diminishes for other values
of $Z$. The energy evolution of the parameter $Z$ is shown in the Table 2
of \cite{ufnel}. Here, in the Table, we show the energy evolution of both $Z$
and $G(s,0)$ for $pp$ and $p\bar p$ scattering.
\medskip
\begin{table}
\medskip
Table.  $\;\;$ The energy behavior of $Z$ and $G(s,0)$.
\medskip

    \begin{tabular}{|l|l|l|l|l|l|l|l|l|l|l|l}
        \hline
$\sqrt s$, GeV&2.70&4.11&4.74&7.62&13.8&62.5&546&1800&7000\\ \hline
Z             &0.64&1.02&1.09&1.34&1.45&1.50&1.20&1.08&1.00 \\  
$G(s,0)$     &0.68&1.00&0.993&0.94&0.904&0.89&0.97&0.995&1.00 \\  \hline
   
\end{tabular}
\end{table}

The function $G(s,b)$ in Eq. (\ref{ge}) has the maximum at $b_m^2=-2B\ln Z$
with full absorption $G(b_m)=1$. Its position depends both on $B$ and $Z$.
Note, that, for $Z>1$, one gets $G(s,b)<1$
at any physical $b$ with the largest value reached at $b=0$ because the maximum
appears at non-physical values of $b$. The disk is
semi-transparent. At $Z=1$, the maximum is positioned exactly at $b=0$, and
$G(s,0)=1$. The disk becomes black in the center. At $Z<1$, the maximum shifts
to positive physical impact parameters. The dip is formed at the center. It
becomes deeper at smaller $Z$. The limiting value $Z=0.5$ is considered in
more details below.

The maximum absorption in central collisions $G(s,0)=1$ is reached at the 
critical point $Z=1$ which is the case at $\sqrt s=7$ TeV considered first. 
Moreover, the strongly absorptive core of the interaction region grows in size
as we see from expansion of Eq. (\ref{ge}) at small impact parameters:
\begin{equation}
G(s,b)= \frac {1}{Z^2}[2Z-1-\frac {b^2}{B}(Z-1)-\frac {b^4}{4B^2}(2-Z)].
\label{gb}
\end{equation}
The second term vanishes at $Z=1$, and $G(b)$ develops a plateau which extends
to quite large values of $b$ about 0.4 - 0.5 fm. Even larger values of $b$ 
are necessary for the third term to play any role at 7 TeV where 
$B\approx 20$ GeV$^{-2}$. The structure of the interaction region with a 
central core is also supported by direct computation \cite{dnec} using the 
experimental data of the TOTEM collaboration \cite{totem1, totem2} about
the differential cross section in the region of $\vert t\vert \leq 2.5$ GeV$^2$.
The results of analytical calculations and the computation practically coincide
(see Fig. 1 in \cite{ads}). It was also shown in \cite{ads} that this 
two-component structure is well fitted
by the expression with the abrupt (Heaviside-like) change of the exponential.
The diffraction cone contributes mostly to $G(s,b)$. Therefore, the 
large-$\vert t\vert $ elastic scattering can not serve as an effective trigger
of the black core even though some models were proposed (see, e.g., 
\cite{kop, kop1, isl, isl1}) which try to elaborate some predictions. 

Inelastic exclusive processes can be effectively used for this purpose. 
One needs such triggers which enhance the contribution due to the central black 
core. Following the suggestions of \cite{fsw, fsw1}, it becomes possible 
\cite{ads} to study the details of the central core using the experimental 
data of CMS collaboration at 7 TeV about inelastic collisions with high 
multiplicity triggered by the jet production \cite{cms} as well as some other 
related data. Separating the core contribution with the help of these triggers, 
one comes to the important conclusion that the simple increase of the 
geometrical overlap area of the colliding protons does not account for
properties of jet production at very high multiplicities. It looks as if the 
parton (gluon) density must strongly increase in central collisions and
rare configurations of the partonic structure of protons are involved.

It is interesting that the positivity of $G(s,b)$ imposes some limits on the
relative role of $B$ and $\sigma _t$. Namely, it follows from (\ref{gZ}) that
\begin{equation}
2Z=\frac {8\pi B}{\sigma _t}\geq 1.
\label{Bsig}
\end{equation}
This relation implies that the slope $B$ should increase asymptotically at
least as strong as the total cross section $\sigma _t$. This inequality must be
fulfilled even at intermediate energies.

It is usually stated that the equality $2Z=8\pi B/\sigma _t=1$ corresponds to 
the black disk limit with equal elastic and inelastic cross sections 
$\sigma _{el}=\sigma _{in}=0.5\sigma _t $. However, one sees that $G(s,b=0)=0$, 
i.e. the interaction 
region is completely transparent in central collisions. This paradox is 
resolved if we write the inelastic profile of the interaction region using
Eq. (\ref{ge}). At $Z=0.5$ it looks like
\begin{equation}
G(s,b)= 4[e^{-\frac {b^2}{2B}}-e^{-\frac {b^2}{B}}].
\label{0.5}
\end{equation}
Recalling that $B=R^2/4$, we see that one should rename the black disk as
a black torus (or a black ring)
with full absorption $G(s,b_m)=1$ at the impact parameter 
$b_m=R\sqrt {0.5\ln 2}\approx 0.59R$, complete transparency at $b=0$ and rather
large half-width about 0.7R. Thus, the evolution to values of $Z$ smaller than 
1 at higher energies (if this happens in view of energy tendency of $Z$ shown
in the Table) would imply quite special transition from the two-scale
features at the LHC to torus-like configurations of the interaction region.
Its implications for inelastic processes are to be guessed and studied.

In principle, the positivity of the inelastic cross section
\begin{equation}
\sigma _{in}=\frac {\pi B}{Z^2}(4Z-1)\geq 0
\label{posin}
\end{equation}
admits the value of $Z$ as small as 0.25 which corresponds to 
$\sigma _{el}=\sigma _t$ and $\sigma _{in}=0$. However, this possibility 
looks unphysical and has no interpretation in terms of eikonal (blackness). 

Another consequence of Eq. (\ref{gZ}) follows from study of energy evolution of 
$G(s,0)$ shown in the Table. In connection with torus-like structure, it is 
interesting to point out the value of $Z=0.64$ or $G(s,0)=0.68$ at 
$\sqrt s=2.70$ GeV and maximum 1 at $b_m^2=4B\ln 2$. One also notices that, in 
the energy interval 4 GeV$<\sqrt s<8$ GeV, the values of $Z$ are slightly larger 
than 1 so that the values of $G(s,0)$ are smaller but very close to 1. It looks
as if the interaction region becomes black at the center $b=0$ but at higher
energies up to ISR loses this property trying to restore it at the LHC. This 
fact asks for further studies in the energy interval 4 GeV$<\sqrt s<8$ GeV 
especially in view of proposed experiments in Protvino.
The dark core must be smaller there than at LHC because of smaller values of 
$B$. Moreover, the contribution due to the real part of the amplitude is larger 
at these energies as well as larger $\vert t\vert $ beyond the diffraction cone 
can be important. One should also notice that $Z$ becomes less than 1 at even 
smaller energies. As is easily shown, that does not pose any problem with
the requirement $G(s,b)\leq 1$ even though, at first sight, some problems could 
arise because the linear in $b^2$ term in Eq. (\ref{gb}) becomes positive.

Now, we come to assumptions used in getting our conclusions. First, the real
part of the amplitude $f$ (or the ratio $\rho $) has been neglected. At LHC,
it is small at $t=0$ and there are theoretical arguments that it is even 
smaller within the diffraction cone. Thus, it looks safely to say that its
contribution to $G(s,b)$ is less than $10^{-2}-10^{-3}$ there. Surely, these
values are within the accuracy of estimates of $Z$ from experimental data.
At lower energies it can become larger (of the order of 0.1) and change the 
conclusions. Second, the differential cross section was
approximated by its diffraction cone expression (\ref{expB}) and no Orear
region was attributed beyond it. Its comparison with fit of TOTEM experimental 
data done in \cite{ads} shows that it also works quite well there with accuracy 
about $10^{-3}$. Nevertheless, at lower energies new analysis should be done.

In this connection, we should mention that the same parameter $Z$ in
combination with $\rho (s,t)$ determines the slope of the differential cross 
section in the Orear region as was shown a long ago \cite{anddre1, anddre2}. 
When $Z=1$, the slope depends only on $\rho $. That allowed to estimate its 
value in the Orear region at 7 TeV \cite{dnec1} which happened to be 
surprisingly large in modulus and negative. No models have yet explained 
this finding.

In conclusion, it is shown that the absorption at the center of the interaction
region of protons is determined by a single energy-dependent parameter 
$Z$. The region of full absorption extends to quite large impact parameters 
if $Z$ tends to 1. This happens at $\sqrt s=7$ TeV where the two-scale 
structure of the interaction region of protons becomes well pronounced. 
That leads to special consequences both for elastic and inelastic processes.
Energy behavior of $Z$ at higher energies is especially important in view
of possible evolution of the geometry of the interaction region. 
\medskip

{\bf Acknowledgments}

\medskip 
I thank S. Denisov and S. Rusakov for comments.
 
I am grateful for support by the RFBR grants 12-02-91504-CERN-a,
14-02-00099  and the RAS-CERN program.

\end{document}